\documentclass[twocolumn]{aastex62}

\usepackage{natbib}
\usepackage{ulem}
\hypersetup{linkcolor=red,citecolor=blue,filecolor=cyan,urlcolor=magenta}
\usepackage{amsmath}

\newcommand{\MO}{$M_\odot$}

\newcommand{\Chandra}{\textit{Chandra}}

\received{}
\revised
\accepted{}
\submitjournal{ApJ}

\shorttitle{Line-of-sight gas sloshing in the cool core of Abell\,907}
\shortauthors{Ueda et al.}

\begin{document}
%\SetRunningHead{Kitayama et al.}{High-resolution Imaging of the SZE}
%\Received{}%{yyyy/mm/dd}
%\Accepted{}%{yyyy/mm/dd}
%\Published{}%{yyyy/mm/dd}

\title{Line-of-sight gas sloshing in the cool core of Abell\,907}

%%% begin:list of authors
% Do NOT capitalize all letters in "textsc".

\correspondingauthor{Shutaro Ueda}
\email{sueda@asiaa.sinica.edu.tw}
\author[0000-0001-6252-7922]{Shutaro Ueda}
\affiliation{Academia Sinica Institute of Astronomy and Astrophysics (ASIAA), No. 1, Section 4, Roosevelt Road, Taipei 10617, Taiwan}

\author{Yuto Ichinohe}
\affiliation{Department of Physics, Rikkyo University, 3-34-1 Nishi-Ikebukuro, Toshima-ku, Tokyo 171-8501, Japan}

\author{Tetsu Kitayama}
\affiliation{Department of Physics, Toho University, 2-2-1 Miyama, Funabashi, Chiba 274-8510, Japan}

\author{Keiichi Umetsu}
\affiliation{Academia Sinica Institute of Astronomy and Astrophysics (ASIAA), No. 1, Section 4, Roosevelt Road, Taipei 10617, Taiwan}

%%% Please use the following style in case that sorting by 
%%% affiliation is impossible. 
%
% \author{%
%   D-Firstname \textsc{D-Familyname}\altaffilmark{1}
%   E-Firstname \textsc{E-Familyname}\altaffilmark{1,2}
%   and
%   F-Firstname \textsc{F-Familyname}\altaffilmark{2}}
% \altaffiltext{1}{Address of Institute}
% \email{ddddd@xxx.xxx.xx.xx}
% \email{eeeee@xxx.xxx.xx.xx}
% \altaffiltext{2}{Address of Institute}

%% `\KeyWords{}' always has to be placed before `\maketitle'.

%\maketitle

\begin{abstract}
We present line-of-sight gas sloshing first found in a cool core in a galaxy cluster. The galaxy cluster Abell\,907 is identified as a relaxed cluster owing to its global X-ray surface brightness taken by the {\it Chandra X-ray Observatory}. The X-ray residual image after removing the global emission of the intracluster medium (ICM), however, shows an arc-like positive excess and a negative excess surrounding the central positive excess in the cluster core, which in turn indicates a disturbance of the ICM. We analyze the X-ray spectra extracted from both regions and find that (1) the ICM temperature and the metal abundance in the positive excess are lower and higher than those in the negative excess, respectively, and (2) the ICM is nearly in pressure equilibrium. We also find a slight redshift difference between the positive and the negative excesses, which corresponds to the velocity shear of $1680^{+1300}_{-920}$\,km\,s$^{-1}$ ($1\sigma$). The X-ray residual image and the ICM properties are consistent with those expected by line-of-sight gas sloshing. Assuming that the gas is moving toward inverse-parallel to each other along the line-of-sight, the shear velocity is expected to be $\sim 800$\,km\,s$^{-1}$.  The velocity field of this level is able to provide non-thermal pressure support by $\sim 34\%$ relative to the thermal one. The total kinetic energy inferred from the shear velocity corresponds to $\sim 30$\,\% of the bolometric luminosity of the sloshing ICM. Abell\,907 is therefore complementary to galaxy clusters in which gas sloshing takes place in the plane of the sky, and is important for understanding gas dynamics driven by sloshing and its influence on the heating to prevent runaway cooling.
\end{abstract}


%%%%%%%%%%%%%%%%%%%%%%%%%%%%%%%%%%%%%%%%%%%%%%%%%%%%%%%%%%%%
\section{Introduction}

Galaxy clusters are dynamically young systems in the universe. A large amount of baryons fills in the gravitational potential well of galaxy clusters that is formed by dark matter. Most of the baryons in galaxy clusters are in the form of X-ray emitting hot gas so-called the intracluster medium (ICM). A cool core composed of dense, cool, and metal rich ICM is often found in the center of galaxy clusters. It is considered that cool cores are formed by radiatively cooling gas, because of the fact that their cooling time estimated by electron density in the core is much shorter than the age of galaxy clusters \citep[e.g.,][]{Peterson06}. The abundance and properties of cool cores in galaxy clusters provide us with a wealth of information about not only the thermal evolution of cosmic baryons, but also the chemical evolution of the ICM around the brightest cluster galaxy (BCG).

The presence of cool cores in galaxy clusters, on the other hand, poses us a challenge regarding the thermal evolution of intracluster baryons. The cooling time of X-ray emitting hot gas is inversely proportional to its electron density squared. Since the gas in the cool core is dense, its cooling time is much shorter than the age of galaxy clusters \citep[e.g.,][]{Peterson06}. This indicates that the gas in the cool core is not able to survive stably for a long time without heating. This is recognized as the cooling problem of the ICM. To maintain the balance between cooling and heating, the feedback of active galactic nuclei (AGN) in the BCG is considered to be a plausible mechanism \citep[see e.g.,][for reviews]{McNamara07, Fabian12}. In fact, X-ray cavities are found in the X-ray surface brightness of a large sample of galaxy clusters \cite[e.g.,][]{Hlavacek-Larrondo15}. Some of them seem to be associated with jet-like radio emissions \citep[e.g.,][]{McNamara05}, which indicates that mechanical energy provided by AGN is a dominant heating source. On the other hand, gas sloshing in the core of galaxy clusters is expected to be another possible heating source. Gas sloshing is induced by minor mergers with a non-zero impact parameter \citep[see][for a review]{Markevitch07}. Numerical simulations show that gas sloshing is able to prevent runaway cooling for at least a few Gyrs \citep[e.g.,][]{Fujita04b, ZuHone10}.

Evidence of gas sloshing is often found in relaxed clusters as a spiral pattern in the residual image of X-ray surface brightness after removing its global profile \citep[e.g.,][]{Churazov03, Clarke04, Blanton11, Owers11, OSullivan12, Canning13, Rossetti13, Ghizzardi14, Sanders14, Ichinohe15, Ueda17, Liu18}. The observed spiral pattern is likely caused by a merger occurring nearly in the plane of the sky, through transport of angular momentum from an infalling galaxy cluster. Since such mergers can occur in all directions, the resulting sloshing planes should be distributed uniformly, and sloshing motions should be found not only in the plane of the sky but also along the line-of-sight (LOS). Although a large fraction of LOS gas sloshing is therefore expected, only a single case in the Virgo cluster has been reported thus far \citep{Roediger11}. This suggests that a large number of LOS gas sloshing systems have been overlooked or misidentified. Recently, \cite{Su17} studied the Fornax cluster focusing on the Kelvin-Helmholtz instability (KHI) and showed that gas sloshing is in fact a reasonable mechanism to suspend runaway cooling. In addition to gas mixing by KHI, the kinetic energy of ICM motions driven by gas sloshing is expected to be an additional heat source in the ICM through its dissipation by turbulence. This contribution, however, strongly depends on the velocity of the sloshing ICM. Only systems of LOS gas motions allow us to directly measure the shear velocity of sloshing motions. Detailed studies of galaxy clusters that are experiencing LOS gas sloshing are, therefore, essential to understand the whole picture of gas sloshing, especially in the context of gas dynamics. To this end, the first step is to identify such a system.

Abell\,907 (hereafter A907) is located at a redshift of $z = 0.167$ \citep{Bohringer07}. The ICM temperature decreases from $\sim 6$\,keV to $\sim 4$\,keV toward the cluster center, and the metal abundance at the cluster center is $\sim 1$\,solar \citep{Vikhlinin05, Vikhlinin06}. These features indicate the presence of a cool core. The X-ray surface brightness of A907 appears to be symmetric, so that this cluster was identified as a relaxed, cool core cluster \citep{Leccardi10, Cavagnolo11}. The total mass enclosed within the virial radius ($\sim 2.3$\,Mpc) is estimated at $\sim 2 \times 10^{15}$\,\MO ~from the Subaru weak-lensing observations \citep{Okabe16}. A907 is therefore classified into one of typical massive, relaxed, cool core clusters. However,   the lack of cavities may suggest that alternative sources of heating are at play. Therefore, A907 is an ideal target to investigate observationally the role of gas sloshing in preventing runaway cooling in the core.  

Throughout the paper, we adopt $\Omega_{\rm m}=0.27$, $\Omega_{\rm \Lambda}=0.73$, and the Hubble constant of $H_{0} = 70\,$km\,s$^{-1}$\,Mpc$^{-1}$ \citep{Komatsu11}. In this cosmology, an angular size of $1''$ corresponds to a physical scale of 2.87\,kpc at the cluster redshift $z=0.167$. Unless stated otherwise, quoted errors correspond to 1$\sigma$.

%%%%%%%%%%%%%%%%%%%%%%%%%%%%%%%%%%%%%%%%%%%%%%%%%%%%%%%%%%%%
\section{Observations and Data Reductions}
\label{sec:obs}

We used the archival X-ray data of A907 taken with the Advanced CCD Imaging Spectrometer \citep[ACIS;][]{Garmire03} on board the {\it Chandra X-ray Observatory}. All three datasets analyzed were taken by the ACIS-I (ObsID: 535, 3185, and 3205). We used the versions of 4.9 and 4.7.8 for \Chandra ~Interactive Analysis of Observations \cite[CIAO;][]{Fruscione06} and the calibration database (CALDB), respectively. After applying {\tt lc\_clean} to the data to exclude the duration of flare, we obtained a net exposure of 106.1\,ksec. We adopted the blank-sky data included in the CALDB as our background data. We extracted the X-ray spectrum of the ICM from interested region of each dataset with {\tt specexctract} in CIAO and combined them after making individual spectrum, response, and ancillary response files for the spectral fitting. We used {\tt XSPEC} version 12.10.0 \citep{Arnaud96} and the atomic database for plasma emission modeling version 3.0.9 in the X-ray spectral analysis, assuming that the ICM is in collisional ionization equilibrium. We also used the abundance table of \cite{Lodders09}.

\section{Analyses and results}

\subsection{X-ray imaging analysis}
\label{sec:image}

\begin{figure*}
 \begin{center}
  \includegraphics[width=8.5cm]{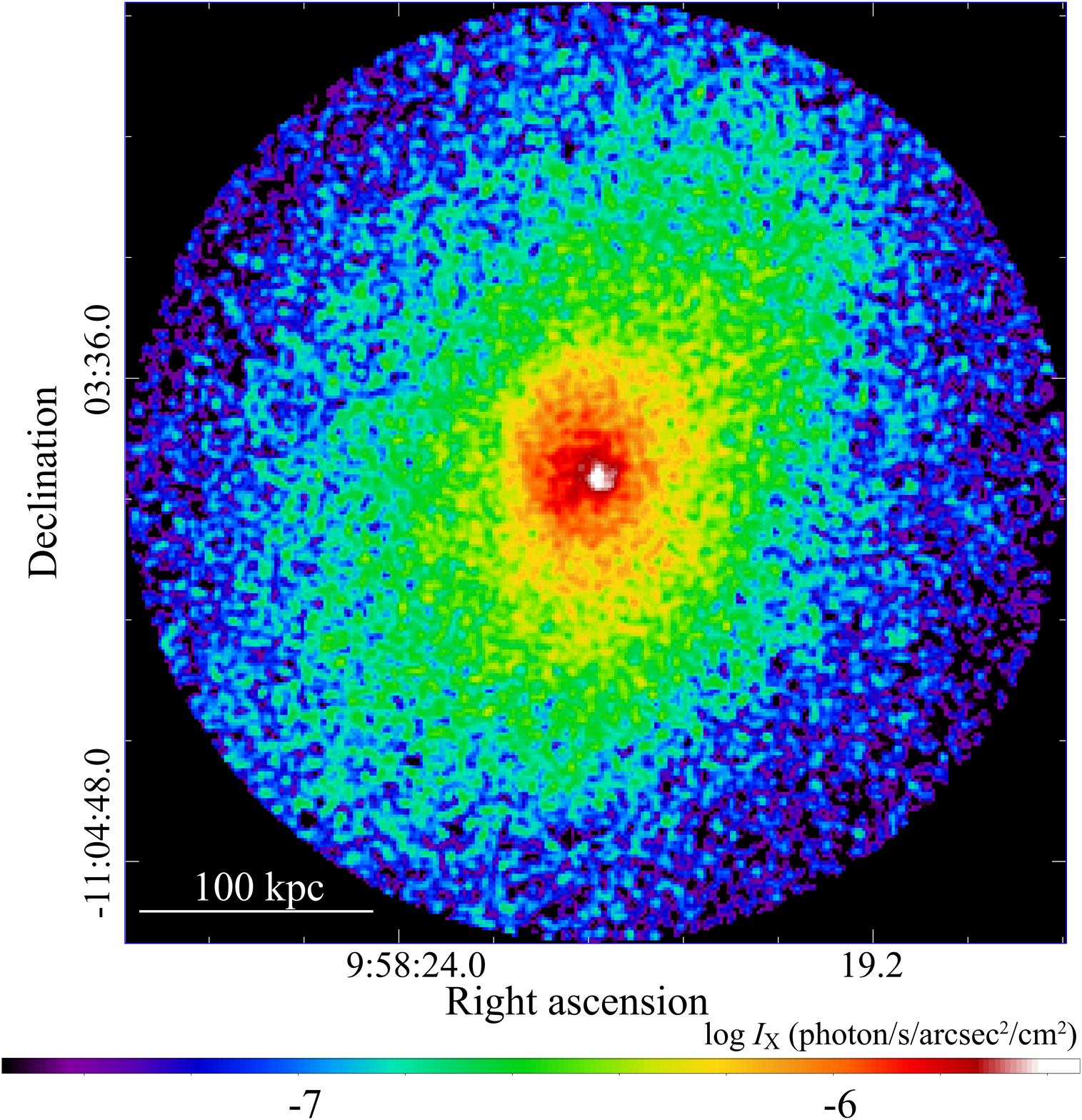}
  \includegraphics[width=8.5cm]{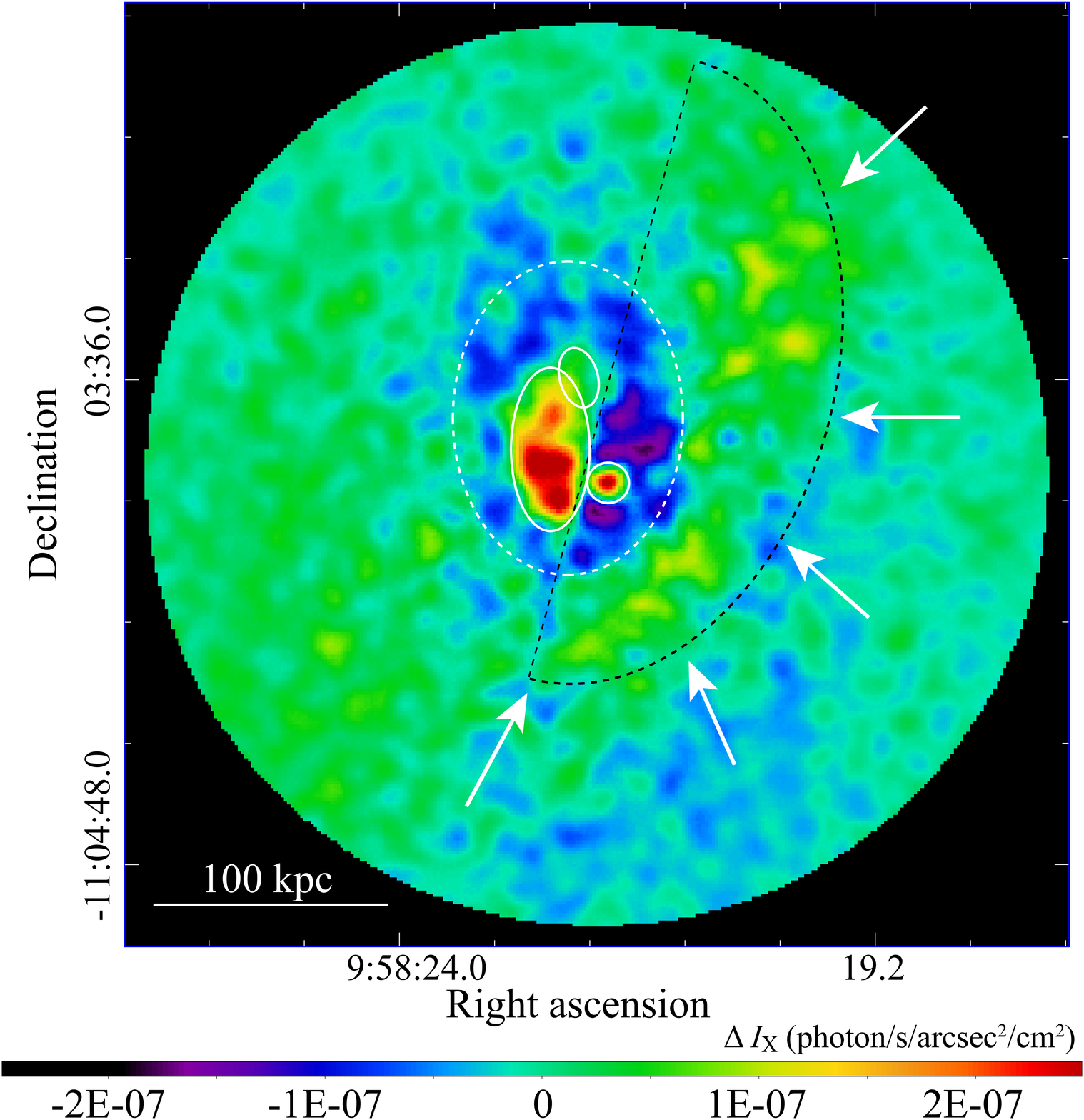} 
 \end{center}
\caption{X-ray surface brightness of A907 (left) and its residual image after removing the mean profile (right).
Left: The X-ray surface brightness in the $0.4 - 7.0$\,keV band is shown as the logarithm value in units of photon\,sec$^{-1}$\,arcsec$^{-2}$\,cm$^{-2}$. This image is smoothed by the Gaussian kernel with $2.3''$ FWHM.
Right: The X-ray residual image of A907 after subtracting the mean profile from the original X-ray surface brightness (the left panel of this figure). Solid, white ellipses show the region to extract the X-ray spectrum of the positive excess region. A dashed, white ellipse is the negative excess region. A black, dashed half ellipse represents the weak positive excess region. The white arrows indicate this region as well.
}
\label{fig:sb}
\end{figure*}

The left panel of Figure~\ref{fig:sb} shows the image of the X-ray surface brightness of A907 in $0.4 - 7.0$\,keV. Since the overall X-ray surface brightness appears to be fairly symmetric, we modeled the mean surface brightness using the concentric-ellipse fitting algorithm of \cite{Ueda17}, by minimizing the variance of the X-ray surface brightness relative to the ellipse model. We found that the axis ratio of this ellipse is $0.70 \pm 0.01$ and its position angle is $165^{\circ} \pm 1^{\circ}$ \footnote{The position angle is measured for the major axis of an ellipse from north ($0^{\circ}$) to east ($90^{\circ}$)}. The position of the center of the ellipse model is located at (RA, Dec) = (9:58:21.979, -11:03:50.065), which is $\sim 0.5''$ offset from that of the BCG. We then subtracted the mean profile from the original X-ray surface brightness. The right panel of Figure~\ref{fig:sb} shows the resultant X-ray residual image of A907. We found a bumpy structure in the X-ray residual image. One is a positive excess with respect to the mean surface brightness profile and the other is a negative excess.

%%%%%%%%%%%%%%%%%%%%%%%%%%%%
%%%%%%%%%%%%%%%%%%%%%%%%%%%%

An arc-like positive excess region is located at a position close to the cluster center. The length of this region is $\sim 75$\,kpc ($26''$) and its width is $\sim 30$\,kpc ($10''$). The angular resolution of \Chandra ~only allows us to detect such small structure. The positive excess region is surrounded by a negative excess region.

In addition, a low-level excess is found in the west side of A907 as indicated with white arrows in the right panel of Figure~\ref{fig:sb}. We call this a weak positive excess region. This region is along the negative excess region and its morphology is arc-like.

\subsection{X-ray spectral analysis}
\label{sec:spec}

\begin{figure}
 \begin{center}
  \includegraphics[width=8.5cm]{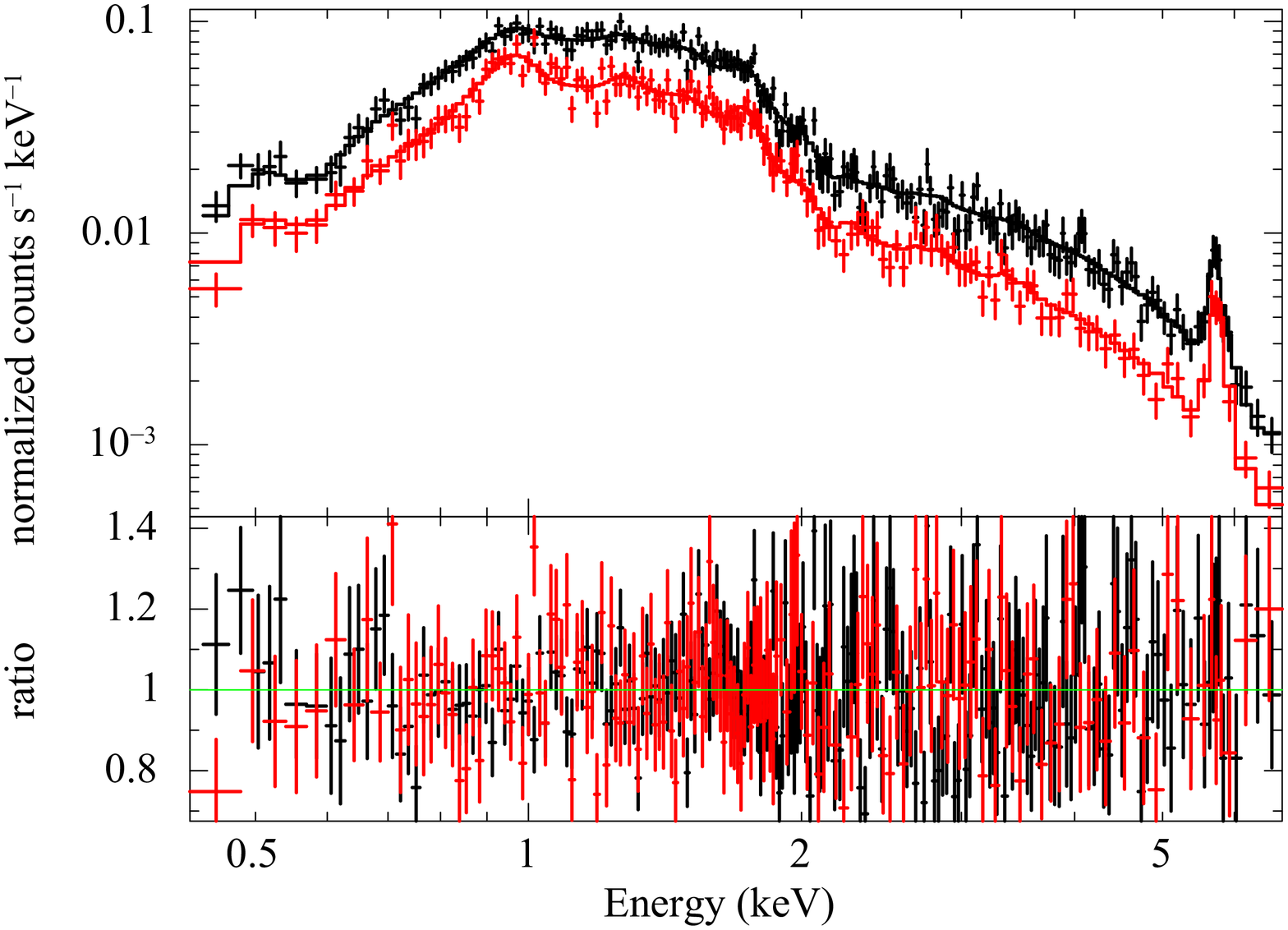}
 \end{center}
\caption{X-ray spectra extracted from the positive excess region (red) and the negative excess region (black) with the best-fit model. The ratios of the data to the model are plotted in the bottom panel.
}
\label{fig:spec}
\end{figure}

To study the origin of the perturbation presented in Section~\ref{sec:image}, we analyzed the X-ray spectra extracted from the positive and the negative excess regions. We show the regions with the positive and the negative excess in the right panel of Figure~\ref{fig:sb} as solid white ellipses and dashed white ellipse, respectively. When we extracted the X-ray spectrum of the negative excess region, we excluded the positive excess region. Figure~\ref{fig:spec} shows the X-ray spectrum of each region and the best-fit model with respect to each X-ray spectrum. The net counts in these regions are shown in Table~\ref{tab:fit}. In the X-ray spectral analysis, we allowed the column density of the Galactic absorption ($N_{\rm H}$) to vary in order to reduce the contamination uncertainty. Table~\ref{tab:fit} shows the best-fit parameters derived by the X-ray spectral analysis of both regions. The electron pressure ($kT \times n_{\rm e}$) and the entropy ($kT \times n_{\rm e}^{-2/3}$) that are readily calculated from the best-fit parameters are also presented in Table~\ref{tab:fit}. We assumed the LOS length of 140\,kpc in this analysis, which is comparable to the projected size of the negative excess region. Note that, although the fitted values of $N_{\rm H}$ are a factor of $\sim 2$ larger than the value inferred from the low resolution 21\,cm line observations by \cite{Kalberla05} (i.e., $5.5 \times 10^{20}$\,cm$^{-2}$), they are consistent with one another within large uncertainties. This is likely due to the uncertainty of contamination calibration and a low spatial resolution map of $N_{\rm H}$. The flux of background data (i.e., cosmic X-ray background and non X-ray background) is two orders of magnitude lower than the source flux, even though the energy band is $2.0 - 7.0$\,keV. The fluctuation of background is therefore negligible.

%%%%%%%%%%%%%%%%%%%%%%%%%%%%
%%%%%%%%%%%%%%%%%%%%%%%%%%%%

%%%%%%%%%%%%%%%%%%%%%%%%%%%%
\begin{table*}[ht]
\begin{center}
\caption{
Best-fit parameters of X-ray spectral analyses in the positive, the negative, and the weak positive excess regions.
}\label{tab:fit}
\begin{tabular}{cccc}
\hline\hline	
Region						&	 Positive excess			&	 Negative excess			&	Weak positive				\\ \hline
Net count ($0.4 - 7.0$\,keV)		&	8372						&	14025					&	16793		\\
Column density $N_{\rm H}$ ($10^{20}$\,cm$^{-2}$)				&	$10.0^{+1.0}_{-1.4}$		&	$9.2 \pm 0.8$		&	$7.6 \pm 1.0$		\\
Temperature (keV)				&	$3.85^{+0.11}_{-0.10}$		&	$4.88^{+0.13}_{-0.14}$		&	$5.42^{+0.27}_{-0.14}$		\\
Abundance ($Z_{\odot}$)			&	$1.17^{+0.11}_{-0.10}$		&	$0.76^{+0.08}_{-0.07}$		&	$0.77^{+0.04}_{-0.05}$		\\
Redshift						&	$0.1652^{+0.0011}_{-0.0028}$	&	$0.1709^{+0.0042}_{-0.0013}$	&	$0.1664 \pm 0.0020$		\\
Density (cm$^{-3}$\,$(L/140\,{\rm kpc})^{-1/2}$)			&	$0.0223 \pm 0.0004$	&	$0.0162 \pm 0.0002$	&	$0.0120^{+0.0002}_{-0.0001}$	\\ 
\hline
\multicolumn{4}{c}{}	\\	\hline
Pressure (keV\,cm$^{-3}$\,$(L/140\,{\rm kpc})^{-1/2}$)	&	$0.086 \pm 0.003$		&	$0.079 \pm 0.002$		&	$0.065^{+0.0033}_{-0.0017}$	\\
Entropy (keV\,cm$^{2}$\,$(L/140\,{\rm kpc})^{1/3}$)		&	$48.6^{+1.5}_{-1.4}$		&	$76.1^{+2.0} _{-2.3}$	&	$103.2^{+5.1}_{-2.6}$		\\
\hline
\end{tabular}
\end{center}
\end{table*}
%%%%%%%%%%%%%%%%%%%%%%%%%%%%

We found that (1) the ICM temperature of the positive excess region is lower ($6\,\sigma$) than that of the negative excess region, (2) the abundance in the positive excess region is higher ($3\,\sigma$) than that in the negative excess region, (3) the electron pressure is comparable to each other ($< 2\,\sigma$), and (4) the entropy in the positive excess region is lower ($10\,\sigma$) than that in the negative excess region. Only statistical errors are considered in calculating the significance levels. In addition, we find a slight redshift difference of $\Delta z = 0.0057^{+0.0043}_{-0.0031}$ between the positive and the negative excess regions, which corresponds to $\Delta v = 1680^{+1310}_{-920}$\,km\,s$^{-1}$. The uncertainty of the detector gain of the ACIS is estimated at $0.3$\,\% \footnote{\url{http://cxc.harvard.edu/cal/summary/Calibration_Status_Report.html}}, which corresponds to $\sim 20$\,eV at 6\,keV. A further study of the gain uncertainty of the ACIS-I was done by \cite{Liu15}, using the background data inside the field of view during the observations of the Bullet cluster. They found a gain uncertainty of $\sim 10$\,eV and no spatial variation on the CCD chip in their observations. The measured redshift offset is about 30\,eV, which is larger than the systematic uncertainty. In addition, both regions are close to each other on the CCD chip, so that this implies that the relative gain uncertainty between the two regions are expected to be canceled, as pointed out by \cite{Liu15}.

We also studied the weak positive excess region by analyzing the X-ray spectrum extracted from this region. In the right panel of Figure~\ref{fig:sb}, a half ellipse (black dashed) represents the weak positive excess region. The X-ray spectrum of the  weak excess region was extracted after excluding the positive and the negative excess regions. The best-fit parameters for this region are summarized in Table~\ref{tab:fit}. We assumed the LOS depth of 140\,kpc in the calculation of the electron density to match with the other excess regions. We find that the ICM temperature of this region is higher than that of the negative excess region and the abundance is lower than that in the positive excess region. In addition, the redshift of this region is consistent with that of A907 (i.e., $z = 0.167$). This result supports that the gain uncertainty during the observations is well-calibrated, so that the redshift offset between the positive and the negative excess regions is reliable. The electron pressure in this region shows a departure from those in the positive and the negative excess regions. The entropy in this region is the largest among the three regions.

\section{Discussion}

\subsection{LOS gas sloshing}
\label{sec:los}

We have measured thermodynamic properties of the ICM in the cluster core through the \Chandra ~observations. We found that the ICM temperature and abundance of the positive excess region are lower and higher than those of the negative excess region, respectively. The electron pressure in both regions suggests that the ICM in the core is nearly in pressure equilibrium. All of the differences and similarities of the ICM are in good agreement with the thermodynamic properties expected by gas sloshing \citep[e,g.,][]{Clarke04, Blanton11, Ichinohe15, Ueda17}: i.e., the ICM in the positive excess region is comprised of cool, dense, and metal rich gas originally in the cool core and is uplifted toward outside the cool core. On the other hand, the ICM in the negative excess region consists of relatively hot, thin, and metal poor gas originated in the outer region and is flowing into the cool core. 

We also find a possible velocity shear between the two regions. This suggests that the two components are moving toward inverse-parallel to each other. This redshift offset appears to be higher than those measured in several galaxy clusters. For example, in the case of Abell\,1835, the upper limit of redshift offset corresponds to $\Delta v < 600$\,km\,s$^{-1}$ \citep{Ueda17}. \cite{Liu18} measured the redshift offset of the cold fronts relative to the cluster average in Abell\,2142 to be $810 \pm 330$\,km\,s$^{-1}$. For the Perseus cluster, \cite{Hitomi18d} reported the velocity gradient with a $100$\,km\,s$^{-1}$ amplitude across the cluster core, which is likely associated with a sloshing motion in the plane of the sky. The relatively high level of the velocity offset in A907 is therefore peculiar. This indicates that the direction of the ICM motion in A907 is different from other galaxy clusters that are experiencing gas sloshing in the plane of the sky. In addition, the morphology of the X-ray residual image of A907 is similar to that expected by LOS gas sloshing in numerical simulations. According to numerical simulations, perturbations in the X-ray surface brightness generated by LOS gas sloshing are not spiral-like but ripple-like \citep[e.g., Figure~5 of][]{Roediger11}. This paper therefore presents for the first time the LOS gas sloshing in the cluster core.

%%%%%%%%%%%%%%%%%%%%%%%%%%%%
\begin{figure*}
 \begin{center}
  \includegraphics[width=8cm]{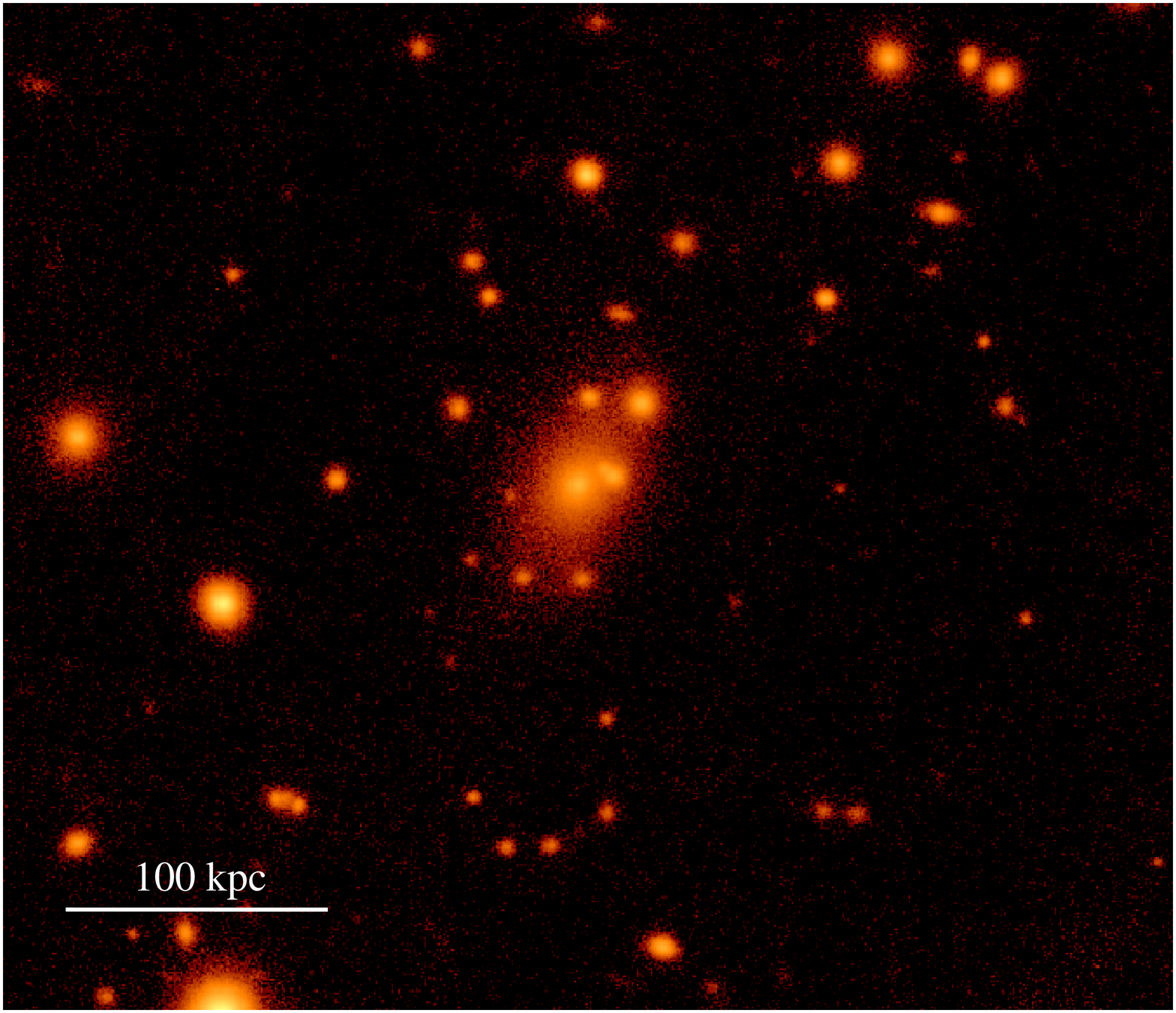}
  \includegraphics[width=8cm]{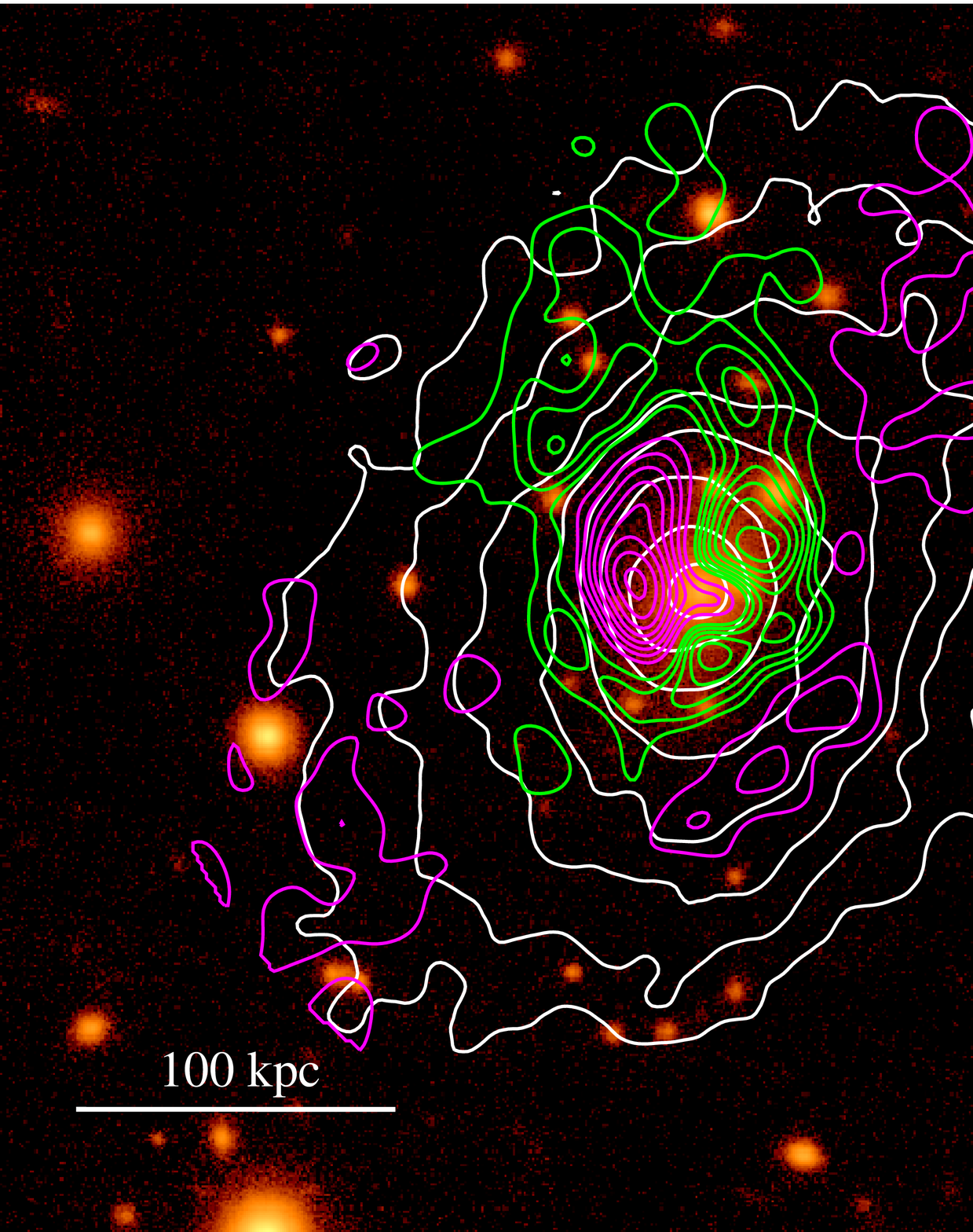}
 \end{center}
\caption{
Left: The optical image of A907 is taken by the Pan-STARRS1 (PS1) $i$-band.
Right: The contours of the X-ray surface brightness (white), the positive excess (magenta), and the negative excess regions (green) are overlaid on the left panel.
}
\label{fig:ps1}
\end{figure*}
%%%%%%%%%%%%%%%%%%%%%%%%%%%%

The left panel of Figure~\ref{fig:ps1} shows the $i$-band optical image of the A907 field observed with the Pan-STARRS1 survey\footnote{\url{https://panstarrs.stsci.edu/}} \citep{Chambers16}. The right panel of Figure~\ref{fig:ps1} presents the contours of the X-ray surface brightness of A907 and the positive excess region overlaid on the optical image. The X-ray centroid coincides with the position of the BCG. The BCG is located at the interface between the positive and the negative excess regions. The positive excess region is extended toward the direction of east from the BCG. This feature is consistent with that expected by LOS gas sloshing in numerical simulations \citep[e.g.,][]{Roediger11}. This similarity supports the scenario of LOS gas sloshing in A907. The quality of this optical image does not allow us to identify the second BCG, which is most likely associated with an infalling subcluster, i.e., the cause of LOS gas sloshing. Deep, high-resolution optical data are needed to search for the second BCG and to study the substructure in the central region of A907 using strong gravitational lensing and the member galaxy distribution.

The thermodynamic property of the ICM in the weak positive excess region is far from that of the positive and the negative excess regions, i.e., the highest ICM temperature and entropy, and also the lowest electron pressure among them. The abundance of this region is consistent with that of the negative excess region. These results imply that the weak positive excess is not generated by the same reason as the central positive and negative excess regions. Since the redshift of this region is consistent with that of A907, no large bulk motion is expected.  According to numerical simulations \citep[e.g.,][]{Roediger11, ZuHone16}, a minor merger also generates a low-level positive fluctuation outside the core, which has a higher temperature and slower motion than those in the central region. These pictures are consistent with those observed in the weak positive excess region. The weak positive excess region may therefore be also caused by a minor merger.

\subsection{Velocity shear in the core}
\label{sec:shear}

The system of LOS gas sloshing in A907 allows us to measure the velocity of the ICM motion induced by gas sloshing. This information is important for understanding gas dynamics as well as a heating source(s) to prevent runaway cooling in the cool core. Some instabilities such as KHI are induced by a velocity shear at the interface of a cold front \citep[e.g.,][]{Roediger12, Roediger13, Su17, Ichinohe17, Ichinohe19}. In addition, the kinetic energy of sloshing ICM is converted into heat through turbulence in the end. A direct measurement of the ICM velocity field in the entire cool core is done by {\it Hitomi} for the Perseus cluster \citep{Hitomi18d}, which is experiencing gas sloshing in the plane of the sky and shows a spiral pattern in the residual X-ray surface brightness \citep[e.g.,][]{Churazov03, Simionescu12, Walker17}. The main component of the velocity field of sloshing motion is still unclear, so that the velocity field inside the core might be underestimated. 

The redshift difference we find indicates a velocity offset of $\Delta v = 1680^{+1310}_{-920}$\,km\,s$^{-1}$. The velocity of each component is then inferred to be $\Delta v / 2$, i.e., $\sim 800$\,km\,s$^{-1}$. This inferred velocity is in good agreement with that expected by gas sloshing in numerical simulations \citep[e.g.,][]{Ascasibar06}. On the other hand, the adiabatic sound speed in the core of A907 is expected to be $\sim 1030$\,km\,s$^{-1}$ at 4\,keV, which is larger than the inferred velocity. In addition, the ICM in both regions is nearly in pressure equilibrium. These are consistent with the picture that the nature of gas sloshing is sub-sonic and in pressure equilibrium \citep[][]{Ueda18}. 

In addition to the thermal pressure, this shear velocity can provide additional, non-thermal pressure support for the ICM. The ratio of the non-thermal to thermal energy density is estimated as $\epsilon_{\rm non} / \epsilon_{\rm therm} = (\gamma / 3) {\cal M}^2$, where $\epsilon_{\rm non}$ is a non-thermal energy density, $\epsilon_{\rm therm}$ is a thermal energy density, $\gamma$ is the adiabatic index of $\gamma = 5/3$, and $\cal M$ is the Mach number, respectively. The ratio is then $\epsilon_{\rm non} / \epsilon_{\rm therm} = 0.34^{+0.73}_{-0.27}$. Note that this value is derived from the shear velocity only between the positive and the negative excess regions. The ratio in the entire core of A907 might be smaller than the observed ratio.In the case of the Perseus cluster, \cite{Hitomi18d} directly measured the non-thermal to thermal ratio in the core to be $\sim 0.02 - 0.07$ using the velocity of the bulk motion along the LOS and assuming isotropic turbulence recovered by the LOS velocity dispersion. Their estimate {\it did not} include the contribution of bulk motion of sloshing in the plane of the sky. Our result indicates that the contribution of bulk motion along the direction of gas sloshing is essential for the contribution of non-thermal pressure. 

\cite{Molnar10} showed using high-resolution cosmological simulations that the fraction of non-thermal pressure is $20 - 45$\,\% in the core of high-mass relaxed clusters, whose virial mass range is in $(1 - 2) \times 10^{15}$\,\MO. This is in good agreement with the virial mass of A907 \citep[$\sim 2 \times 10^{15}$\,\MO,][]{Okabe16}. To compare with their results, we recalculate the non-thermal fraction using their definition, i.e., $p_{\rm nth} / (p_{\rm nth} + p_{\rm therm})$, where $p_{\rm nth}$ is non-thermal pressure and $p_{\rm therm}$ is thermal pressure. The fraction is then $\sim 25$\,\%, which is consistent with that suggested by \cite{Molnar10}. On the other hand, another cosmological simulations suggested that the non-thermal pressure support in the cluster core is about 10\,\% in relaxed clusters \citep[e.g.,][]{Lau09, Nelson14}. Note that their mass range is different from that of \cite{Molnar10}. This level is consistent with that in the Perseus cluster \citep{Hitomi18d}. To address this tension, it is crucial to reveal a 3D gas motions in cluster cores for a large sample of systems, comprehending the nature of galaxy clusters. Since gas sloshing is associated with a minor merger and its frequency depends on the structure formation of galaxy clusters, understanding gas dynamics driven by gas sloshing plays a key role in studying the evolution of the ICM in the cool core and in measuring the total mass of galaxy clusters. A907 is therefore one of important sources to reveal the 3D motion of the ICM in the cluster core.

\subsection{Gas dynamics driven by gas sloshing}
\label{sec:dy}

It is expected that the kinetic energy inferred from the sloshing motion of the ICM is finally converted into heat. If the contribution of such kinetic energy is significant enough to prevent runaway cooling in the cool core for a certain period, the heating by gas sloshing should be complementary to the AGN feedback. 

First, we estimate the total mass in the positive and the negative excess regions, assuming the LOS depth of 140\,kpc, the mean molecular weight of 0.6, and the obtained electron number density. The total mass of the ICM in both regions is then $\sim 1.1 \times 10^{11}$\,\MO ~and $\sim 2.7 \times 10^{11}$\,\MO, respectively. In this case, the total kinetic energy of each region is estimated to be $\sim 0.7 \times 10^{60} \times (v / 800\,{\rm km\,s^{-1}})^2$\,erg and $\sim 1.7 \times 10^{60} \times (v / 800\,{\rm km\,s^{-1}})^2$\,erg, respectively, where $v$ is the LOS velocity of the ICM motion. Although the ICM is likely moving not only along the LOS but also in the plane of the sky, it is hard to measure the velocity component on the plane of the sky. We therefore focus on the LOS velocity in this estimate. When we assume that such kinetic energy is released during 1\,Gyr constantly, the estimated power in total taking the statistical error into account is $7.6^{+19.4}_{-3.2} \times 10^{43} \times (t/1\,{\rm Gyr})^{-1}$\,erg\,s$^{-1}$, where $t$ is a time-scale of release of kinetic energy. 

Next, we assume the bolometric luminosity of the ICM as the absorption-corrected X-ray luminosity in the $0.01 - 100$\,keV band. They are then $(1.5 \pm 0.2) \times 10^{44}$\,erg\,s$^{-1}$ and $(8.6 \pm 0.2) \times 10^{43}$\,erg\,s$^{-1}$, respectively. The total bolometric luminosity in the sloshing core is thus $(2.4 \pm 0.2) \times 10^{44}$\,erg\,s$^{-1}$. The ratio of the estimated power to the bolometric luminosity is $0.32^{+0.81}_{-0.14}$, which indicates, albeit with large statistical errors, that gas sloshing may play a significant role in heating the ICM. Note that this is inferred from the LOS velocity alone, so that this value is a lower limit of the power. The remaining required power is however large, which means that an additional heating source(s), such as the AGN feedback is needed, or the time-scale of energy release should be less than $\sim 0.3$\,Gyr to keep the complete balance between cooling and heating solely by the total kinetic energy. Further numerical simulations and observations are needed to estimate the time-scale and efficiency of heating. i.e., to study whether or not gas sloshing can be a heating source.

\section{Conclusions}

Using the archival data of the {\it Chandra X-ray Observatory}, we have presented the first detection of LOS gas sloshing in the cool core, which appears as a characteristic perturbation in the X-ray surface brightness of the core of A907. The conclusions of this paper are summarized as follows.

\begin{enumerate}

\item In the residual image of the X-ray surface brightness after subtracting the mean profile of A907, we find the positive and the negative excess regions. The positive excess region is surrounded by the negative excess region. The height and width of the positive excess region are $\sim 75$\,kpc ($26''$) and $\sim 30$\,kpc ($10''$), respectively.

\item We analyze the X-ray spectra extracted from the positive and the negative excess regions and find that the ICM thermodynamic properties of both regions are in good agreement with those expected by gas sloshing. We also find a slight redshift difference between both regions, which corresponds to the velocity shear of $\Delta v = 1680^{+1310}_{-920}$\,km\,s$^{-1}$. This level of velocity offset is the highest among the previously measured systems of gas sloshing. When the ICM in both regions is moving toward inverse-parallel to each other, the shear velocity is expected to be $\sim 800$\,km\,s$^{-1}$. 

\item All of the observational results indicate that A907 is experiencing gas sloshing along the LOS. LOS gas sloshing in the cluster core is found for the first time. A907 is important for understanding gas dynamics driven by gas sloshing and is complementary to galaxy clusters that have gas sloshing in the plane of the sky.

\item The observed shear velocity suggests that the ratio of the non-thermal to thermal energy density is about 0.3. A907 is identified as one of typical massive, relaxed, cool core clusters, while our results indicate that the non-thermal fraction of this level is not negligible at least in the cluster core. This indicates that revealing 3D bulk motion induced by gas sloshing is crucial to measure non-thermal pressure. 

\item The total kinetic energy inferred from the shear velocity and the ICM mass allows us to estimate the total power released in gas sloshing. The total estimated power is $\sim 7.6 \times 10^{43} \times (t/1\,{\rm Gyr})^{-1}$\,erg\,s$^{-1}$ if the energy release rate is constant during gas sloshing. This value corresponds to $\sim 30$\,\% of the bolometric ICM luminosity, which indicates that the contribution of gas sloshing to heat the ICM is non-negligible. To keep the complete balance between cooling and heating by only the total kinetic energy of gas sloshing, the time-scale of energy release should be less than $\sim 0.3$\,Gyr.

\end{enumerate}

\acknowledgments
We are grateful to the anonymous referee for helpful suggestions and comments.
The scientific results of this paper are based in part on data obtained from the Chandra Data Archive: ObsID 535, 3185, and 3205.
This work is supported in part by the Ministry of Science and Technology of Taiwan (grant MOST 106-2628-M-001-003-MY3) and by Academia Sinica (grant AS-IA-107-M01). 
This work is also supported by the Grants-in-Aid for Scientific Research by the Japan Society for the Promotion of Science with KAKENHI Grant Numbers JP18K03704 (T.K.).

\vspace{5mm}
\facilities{CXO}

% See the manual for the detail.
%%%

\bibliographystyle{apj}
\bibliography{00_BibTeX_library}

\end{document}